# Fuzzy overlapping communities in networks

Steve Gregory

Department of Computer Science, University of Bristol, Bristol BS8 1UB, England

**Abstract.** Networks commonly exhibit a community structure, whereby groups of vertices are more densely connected to each other than to other vertices. Often these communities overlap, such that each vertex may occur in more than one community. However, two distinct types of overlapping are possible: crisp (where each vertex belongs fully to each community of which it is a member) and fuzzy (where each vertex belongs to each community to a different extent). We investigate the effects of the fuzziness of community overlap. We find that it has a strong effect on the performance of community detection methods: some algorithms perform better with fuzzy overlapping while others favour crisp overlapping. We also evaluate the performance of some algorithms that recover the belonging coefficients when the overlap is fuzzy. Finally, we investigate whether real networks contain fuzzy or crisp overlapping.

## 1. Introduction

Networks are a natural representation of various kinds of complex system, in society, biology, and other fields. Although the study of networks is not new, the amount of network data has proliferated in recent years, thanks to developments in computing and communications technology. As the number and size of network datasets has increased, so too has interest in computational techniques that help us to understand the properties of networks.

A key property of many networks is their community structure: the tendency for vertices to be gathered into distinct groups, or *communities*, such that edges between vertices in the same community are dense but intercommunity edges are sparse. Identifying communities can allow us to understand attributes of vertices from network topology alone. For example, the vertices in a community may be related in some way. The automatic discovery of network communities can also help reveal the coarse-grained structure of networks which are too large for humans to make sense of at the level of individual vertices.

Numerous community detection algorithms have been developed, using a variety of techniques: removal of high-betweenness edges [1], modularity optimization [2, 3], detection of dense subgraphs [4], statistical inference [5], and many more. Even a brief description of these algorithms is beyond the scope of this paper. The interested reader is referred instead to Fortunato's excellent, comprehensive survey [6] of community detection.

Unfortunately there is no generally accepted definition of *community* [6, 7]; each algorithm makes different assumptions that are consistent with the intuitive concept. Most assume that a network contains a flat set of disjoint communities. This makes sense for many networks: for example, most employees work for a single employer. Some algorithms [4, 8–18] allow communities to overlap. This may be more realistic: for example, researchers sometimes belong to more than one research group. Yet other algorithms [11, 19, 20] can find a hierarchy of nested communities, such as a department that comprises a number of research groups. Hierarchy is a significant special case of overlap.

In the context of (non-hierarchical) overlapping communities, it is possible to distinguish between two forms of overlap. With *non-fuzzy* or *crisp* overlapping, each individual (network vertex) belongs to one or more communities with equal strength: an individual either belongs to a community or it does not. With *fuzzy* overlapping, each individual may also belong to more than one community but the strength of its membership to each community can vary. The strength of membership of vertex *v* to community *c* is usually expressed as a *belonging coefficient*, $\alpha_{vc}$: a real number between 0 and 1 such that, for every *v*,

$$\sum_c \alpha_{vc} = 1. \tag{1}$$

Belonging coefficients describe how a given vertex is distributed between communities. Occasionally *association levels* are used instead [16]: these measure the relative contribution of each vertex to a given community, summing to 1 for all vertices in that community. Another related concept is the participation coefficient [21] of a vertex, which measures how well distributed are the vertex's neighbours among different communities.

Examples of both crisp and fuzzy overlapping can readily be found in real networks. For example, in a social network of the type typified by Facebook, a person often belongs to many communities of different types: colleagues, former colleagues, relatives, etc. This is an example of crisp overlapping. Conversely, in a collaboration network of researchers, the overlapping may be fuzzy because a researcher who belongs to several communities cannot be fully involved with all of them, as a result of limited time and resources. Fuzzy and crisp overlapping can also be found in biological networks [21] and other types of network.

Most of the work that has been done to date on detecting and evaluating overlapping communities has assumed one form of overlapping (fuzzy or crisp) and has not considered the alternative. Several questions remain unanswered:

1. Does the type of overlapping in a network affect the ability of an algorithm to detect overlapping communities?
2. How can "fuzzy" algorithms (those that produce a fuzzy partition[1]) be compared with "crisp" algorithms (which produce a crisp partition)?
3. Can a crisp algorithm be modified to produce a fuzzy partition, and vice versa?
4. Do real networks contain fuzzy or crisp overlapping?

This paper seeks to answer these questions. Section 2 surveys some of the algorithms proposed to detect overlapping communities and the measures proposed to evaluate them. In section 3 we consider the similarities and differences between fuzzy and crisp overlapping in networks and between fuzzy and crisp algorithms. Section 4 presents results of experiments on both of the issues discussed in section 3, on synthetic and real networks. Conclusions appear in section 5.

**2. Background**

*2.1. Overlapping community detection algorithms*

Most algorithms for detecting overlapping communities are crisp, in the sense that they produce a crisp partition (containing no belonging coefficients). In one of the first such algorithms, Baumes *et al* [8] proposed a two-phase method whereby a network is first broken into a number of disjoint "seed" communities and then each community is grown by adding and removing adjacent vertices until its "density" is maximized. This density function (not to be confused with the common concept of *graph density*) is a function of each community, and so is quite cheap to compute. The algorithm relies on finding a *local* maximum of density; the global maximum corresponds to the trivial case where the network contains a single community.

The LFM method of Lancichinetti *et al* [11] is very similar to that of Baumes *et al*: it expands seed communities in the same way, to find a local maximum of a fitness function similar to that of [8]. The main difference is that a seed community is simply any vertex that is not yet assigned to any community. Lee *et al* [13] recently developed this idea further by using maximal cliques, instead of individual vertices, as seed communities. Their "greedy clique expansion" (GCE) algorithm has the important advantage that it can detect a much higher degree of overlap. EAGLE [18] is another algorithm that uses maximal cliques to find overlapping communities.

---
[1] A partition is often called a *cover* when its communities overlap, and a *fuzzy partition* or *fuzzy cover* if the overlap is fuzzy, but we use the term partition throughout this paper.

Palla *et al* [4] define a community as a set of *k*-cliques each of which shares at least *k*–1 vertices with another *k*-clique in the set. CFinder is an algorithm to locate such communities, which may overlap, for any given *k*.

CONGA [9, 10] and COPRA [14] are both "overlapping" versions of existing disjoint community detection algorithms. CONGA extends the algorithm of Girvan and Newman [1] with the ability to split a vertex into two vertices, possibly repeatedly, during the divisive clustering process; the multiple copies of a vertex can be placed in different communities, resulting in overlap. COPRA extends the label propagation algorithm [22] to allow overlap by retaining multiple community labels on each vertex.

Fewer fuzzy methods (those that produce fuzzy partitions) have been proposed. Nepusz *et al* [15] cast the task as a non-linear constrained optimization problem and describe a quadratic-time algorithm to solve it. Zhang *et al* [12] convert a network to ($k$–1)-dimensional Euclidean space and use the fuzzy *c*-means algorithm to detect up to *k* communities. Psorakis *et al* [17] present a method based on Bayesian non-negative matrix factorization (NMF). Finally, FOG [16] is a stochastic framework and algorithm for clustering "link data", which includes networks as a special case, into fuzzy communities. FOG differs from the other fuzzy algorithms by computing association levels instead of belonging coefficients.

*2.2. Overlapping modularity measures*

The modularity measure was introduced in [23] to measure the quality of a disjoint partition of a network. Modularity is defined in equation (2) and (equivalently) equation (3):

$$Q = \frac{1}{2m} \sum_{i,j \in V} \left[ A_{ij} - \frac{k_i k_j}{2m} \right] \delta_{c(i),c(j)}, \tag{2}$$

$$Q = \frac{1}{2m} \sum_{c \in C} \sum_{i,j \in V_c} \left[ A_{ij} - \frac{k_i k_j}{2m} \right]. \tag{3}$$

Here, *V* is the set of vertices in the network, *C* is the partition (a set of communities), [$A_{ij}$] is the adjacency matrix, $k_i$ is the degree of vertex *i*, *m* is the number of edges in the network, $c(i)$ is the community to which vertex *i* belongs, $V_c$ is the set of vertices in community *c*, and $\delta$ is the Kronecker delta.

In equation (3), the first term ($\sum A_{ij}/2m$) is the fraction of edges that fall within communities and the second term ($\sum k_i k_j /4m^2$) is the fraction that would be expected according to the standard null model (the "configuration model"), in which the degree sequence of the network is preserved.

Modularity is not defined when communities overlap, but a few authors have proposed extensions of modularity to networks with overlapping communities. Most of these assume fuzzy overlapping. Nepusz *et al* [15] extend modularity by replacing the Kronecker delta in equation (2), which indicates whether two vertices are in the same community, by a fuzzy similarity measure: $s_{ij}$ is the sum of the products of the belonging coefficients of *i* and *j* in communities to which they both belong:

$$Q = \frac{1}{2m} \sum_{i,j \in V} \left[ A_{ij} - \frac{k_i k_j}{2m} \right] s_{ij} \text{ where } s_{ij} = \sum_{c \in C} \alpha_{ic} \alpha_{jc}. \tag{4}$$

We shall call $s_{ij}$ the *comembership* of *i* and *j*: it measures the extent to which they belong to the same communities.

Shen *et al* [24], apparently unaware of [15], proposed an identical measure:

$$Q = \frac{1}{2m} \sum_{c \in C} \sum_{i,j \in V} \left[ A_{ij} - \frac{k_i k_j}{2m} \right] \alpha_{ic} \alpha_{jc}. \tag{5}$$

The modularity function of Zhang *et al* [12] is more complicated; the main difference is that it measures the similarity of two vertices as the average, not the product, of their belonging coefficients:

$$Q = \frac{1}{2m} \sum_{c \in C} \left[ T_c - \frac{(T_c + U_c)^2}{2m} \right] \tag{6}$$

where $T_c = \sum_{i,j \in V_c} \frac{A_{ij}(\alpha_{ic} + \alpha_{jc})}{2}$ and $U_c = \sum_{i \in V_c, j \in V \setminus V_c} \frac{A_{ij}(\alpha_{ic} + (1 - \alpha_{jc}))}{2}$.

Nicosia *et al* [25] propose the following measure, expressed in terms of a function $F$:

$$Q = \frac{1}{2m} \sum_{c \in C} \sum_{i,j \in V} \left[ A_{ij} F(\alpha_{ic}, \alpha_{jc}) - \frac{k_i k_j \left( \sum_{v \in V} F(\alpha_{vc}, \alpha_{jc}) \right) \left( \sum_{v \in V} F(\alpha_{ic}, \alpha_{vc}) \right)}{2mn^2} \right], \tag{7}$$

where $F(\alpha_{ic}, \alpha_{jc})$ could be defined as a product $\alpha_{ic}\alpha_{jc}$, an average $(\alpha_{ic}+\alpha_{jc})/2$, a maximum $max(\alpha_{ic},\alpha_{jc})$, or any other suitable function.

All of the above measures assume fuzzy overlapping. The only modularity function designed for *crisp* overlapping is one proposed by Lázár *et al* [26]. It defines the modularity as the average of $M_c$ over all communities $c$:

$$M = \frac{1}{|C|} \sum_{c \in C} M_c. \tag{8}$$

The modularity $M_c$ of community $c$ is defined as:

$$M_c = \frac{m_c}{|V_c|(|V_c| - 1)/2} \cdot \frac{1}{|V_c|} \sum_{i \in V_c} \frac{\sum_{j \in V_c, i \neq j} A_{ij} - \sum_{j \notin V_c} A_{ij}}{k_i s_i}, \tag{9}$$

where $m_c$ is the number of edges in community $c$ and $s_i$ is the number of communities to which vertex $i$ belongs. The first factor in equation (9) is the edge density of community $c$, and the second factor measures the difference between the number of intercommunity edges and the number of intracommunity edges, to vertices in $c$, suitably normalized.

### 2.3. Partition comparison measures

An indispensable tool for any clustering task (not only of network data) is a measure to assess the similarity between a pair of partitions. This is often used to measure the quality of a "found" partition when the "real" partition is known, and to measure the stability of a partition over time or when different community detection algorithms are used.

For disjoint partitions there are two widely used measures, each of which maps a pair of partitions to a real number between 0 (meaning that the partitions are totally different) and 1 (meaning they are identical). One is the Normalized Mutual Information measure [27]. The other is the Adjusted Rand Index [28], defined as:

$$r(C_1, C_2) = \frac{r_u(C_1, C_2) - r_e(C_1, C_2)}{1 - r_e(C_1, C_2)}. \tag{10}$$

$r_u(C_1, C_2)$ (the unadjusted Rand Index) is the fraction of pairs that belong to the same community or belong to different communities in *both* partitions $C_1$ and $C_2$:

$$r_u(C_1, C_2) = \frac{|s(C_1) \cap s(C_2)| + |d(C_1) \cap d(C_2)|}{N}, \tag{11}$$

where $s(C)$ is the set of pairs of items that belong to the same community in $C$, $d(C)$ is the set of pairs of items in different communities in $C$, and $N$ ($= n(n-1)/2$) is the total number of pairs. $r_e(C_1, C_2)$ is the expected value of the same fraction in the null model:

$$r_e(C_1, C_2) = \frac{|s(C_1)||s(C_2)| + |d(C_1)||d(C_2)|}{N^2}. \tag{12}$$

Both of these measures have been extended to handle crisply overlapping communities. The Normalized Mutual Information (NMI) measure was extended by Lancichinetti *et al* [11]. An overlapping version of the Adjusted Rand Index is the Omega Index [29], defined as:

$$o(C_1, C_2) = \frac{o_u(C_1, C_2) - o_e(C_1, C_2)}{1 - o_e(C_1, C_2)}. \tag{13}$$

$o_u(C_1,C_2)$ is the fraction of pairs that occur together in the same number of communities in both partitions:

$$o_u(C_1, C_2) = \frac{1}{N} \sum_j |t_j(C_1) \cap t_j(C_2)|, \tag{14}$$

where $t_j(C)$ is the set of pairs of items that appear together in exactly $j$ communities in partition $C$. $o_e(C_1,C_2)$ is the expected value of this fraction in the null model:

$$o_e(C_1, C_2) = \frac{1}{N^2} \sum_j |t_j(C_1)||t_j(C_2)|. \tag{15}$$

Very few measures have been proposed for comparing fuzzy partitions. As far as we are aware, only one of these can be used to measure the similarity between two arbitrary fuzzy partitions: the Fuzzy Rand Index of Hüllermeier and Rifqi [30]. This can best be explained by first redefining the original (unadjusted) Rand Index:

$$r_u(C_1, C_2) = \frac{s(C_1, C_2)}{N}, \tag{16}$$

where $s(C_1,C_2)$ is the number of pairs that occur in the same community *or* in different communities in both $C_1$ and $C_2$. This can be defined in terms of a function $eq(i,j,C)$ which is 1 or 0 depending on whether $i$ and $j$ appear in the same community in $C$:

$$s(C_1, C_2) = N - \sum_{i,j \in V} |eq(i, j, C_1) - eq(i, j, C_2)|. \tag{17}$$

The expected Rand Index can also be redefined:

$$r_e(C_1, C_2) = \frac{s(C_1)s(C_2) + (N - s(C_1))(N - s(C_2))}{N^2}, \tag{18}$$

where $s(C)$ is the number of pairs that occur in the same community in $C$:

$$s(C) = \sum_{i,j \in V} eq(i, j, C), \tag{19}$$

and the $eq$ function is defined as:

$$eq(i, j, C) = 1 \text{ if } \exists c \in C [i \in c \land j \in c] \text{ else } 0. \tag{20}$$

The Fuzzy Rand Index follows naturally from this: the $eq(i,j,C)$ function is replaced by a fuzzy variant indicating the extent to which $i$ and $j$ occur in the same community in $C$, which is dependent on the belonging coefficients of $i$ and $j$. Hüllermeier and Rifqi [30] suggest defining $eq$ as:

$$eq(i, j, C) = 1 - \frac{1}{2} \sum_{c \in C} |\alpha_{ic} - \alpha_{jc}|, \tag{21}$$

and point out that the Fuzzy Rand Index is a metric if $eq$ is defined thus and certain other conditions hold.

An advantage of the Fuzzy Rand Index and the Omega Index is that they are identical to each other, and to the original Adjusted Rand Index, when there is no overlapping. Conversely, the "overlapping" NMI differs slightly from the original NMI measure [11], and has not been extended to fuzzy overlapping.

**3. Fuzziness in overlapping communities**

*3.1. Fuzziness in networks*
In networks with disjoint communities, it is usually assumed that $p_{ij}$, the probability of an edge $\{i,j\}$, depends upon whether $i$ and $j$ are in the same community. If they are, $p_{ij}$ is $p_{in}$ and otherwise it is $p_{out}$, such that $p_{out} < p_{in}$ (usually $p_{out} \ll p_{in}$). For networks with crisply overlapping communities, a similar assumption is made: $p_{ij}$ depends on the number of communities in which $i$ and $j$ occur together. According to Sawardecker *et al* [31], if $i$ and $j$ occur together in $k$ communities, $p_{ij} = p_k$ where $p_0 < p_1 \leq p_2 \leq \ldots$. Probably the simplest definition of $p_{ij}$ that satisfies this is:

$$p_{ij} = p_1 \; if \; \exists c \in C [i \in c \wedge j \in c] \; else \; p_0 . \tag{22}$$

When overlapping is fuzzy, $p_{ij}$ depends not only on the number of communities in which $i$ and $j$ both appear, but also on their degree of belonging to such communities. We propose the definition:

$$p_{ij} = s_{ij} p_1 + (1 - s_{ij}) p_0 , \tag{23}$$

where $s_{ij}$ is the comembership of $i$ and $j$, as defined in equation (4). In principle, $s_{ij}$ could be defined in other ways, analogously to the $F$ function in equation (7).

There are many other ways in which crisp and fuzzy overlapping can be defined, but for simplicity we will use only these two in this paper. Equation (22) will be used for networks with crisp overlapping (which we call "crisp networks") and equation (23) for networks with fuzzy overlapping ("fuzzy networks").

To discover the effects of the two forms of overlapping, we generate synthetic networks that differ only in the definition of $p_{ij}$ used, other characteristics being the same. The networks are all based on randomly generated partitions with overlapping communities, which for fuzzy networks contain random belonging coefficients. We use these networks in our experiments (section 4) to determine what effect the form of overlapping has on community detection.

*3.2. Fuzziness of algorithms*
Algorithms to detect overlapping communities are either "crisp" or "fuzzy" by design: they produce crisp or fuzzy partitions regardless of the type of overlapping in the network. To compare these algorithms consistently, we propose using a common measure: the Fuzzy Rand Index.

1. To evaluate a fuzzy algorithm on a fuzzy network, we compare the fuzzy partition used to construct the network with the one produced by the algorithm.
2. To evaluate a crisp algorithm on a fuzzy network, we first convert the partition found by the algorithm to a fuzzy form by adding equal belonging coefficients for each community. That is, if vertex $v$ belongs to $K$ communities in the crisp partition, its belonging coefficient is $1/K$ in those communities and zero in other communities, in the fuzzy partition. One would expect this trivial fuzzy partition to be worse than one found by a good fuzzy algorithm, because it contains no information about the belonging coefficients.
3. To evaluate a fuzzy algorithm on a crisp network, we convert the crisp partition used to construct the network to a fuzzy form in the same way, and compare it with the fuzzy partition found by the algorithm.
4. If both the network and the algorithm are crisp, we convert both partitions (the original one and that found by the algorithm) to a fuzzy form and compare them using the Fuzzy Rand Index. In this special case, the partitions could instead be compared by a non-fuzzy measure such as the Omega Index or NMI.

Finally, we describe a simple procedure for obtaining a *non-trivial* fuzzy partition from a crisp one. For each occurrence of vertex $i$ in community $c$, we add a belonging coefficient $\alpha_{ic}$ which equals the number of *i*'s neighbours that occur in *c* divided by the size of *c*, normalized in the usual way. This technique, which we call *MakeFuzzy*, can be used to convert any crisp algorithm to a fuzzy one, which may produce better solutions than the crisp algorithm; we test this hypothesis in our experiments.

## 4. Experiments

### 4.1. Methodology

To experiment with fuzzy and crisp overlapping communities, we have developed a method to generate artificial networks with both types of overlapping, based on the *benchmark* network generator of Lancichinetti *et al* [32], which we shall call the LFR method. The LFR method produces networks that are claimed to possess properties found in real networks, such as heterogeneous distributions of degree and community size. It also allows communities to overlap, though this is not described in [32]. However, it is not directly suitable for our purposes because it does not allow the fuzziness of overlapping to be varied.

The LFR method generates a set of communities, and a network based on them, that satisfy the user's parameters. Some of the parameters specify properties of communities: $N$ (number of vertices), $c_{\min}$ and $c_{\max}$ (minimum and maximum community size), $\tau_2$ (exponent of the power-law distribution of community sizes), $o_m$ (number of communities each "overlapping" vertex belongs to), and $o_n$ (number of "overlapping" vertices: those that are in more than one community).

The other parameters specify properties of the generated network: $\langle k \rangle$ (average degree), $k_{\max}$ (maximum degree), $\mu$ (mixing parameter: each vertex shares a fraction $\mu$ of its edges with vertices in other communities), and $\tau_1$ (exponent of the power-law distribution of vertex degrees).

Our fuzzy network generator (figure 1) produces a set of *fuzzy* communities and a network from the output of the LFR benchmark generator. First, the crisp communities are converted to fuzzy form by adding a random belonging coefficient to each occurrence of each item. These are uniformly distributed: for a vertex *i* that belongs to two communities, *c* and *d*, $\alpha_{ic}$ is drawn from a uniform distribution between 0 and 1 and $\alpha_{id}$ is $1-\alpha_{ic}$. Next, a new network is constructed from these fuzzy communities, using either the fuzzy or crisp formula of section 3.1. In equations (22) and (23), $p_0$ and $p_1$ are chosen so as to preserve the specified average degree ($\langle k \rangle$) and mixing parameter ($\mu$) in the generated network. The final network then satisfies all of the original parameters with the exception of the degree distribution ($k_{\max}$ and $\tau_1$).

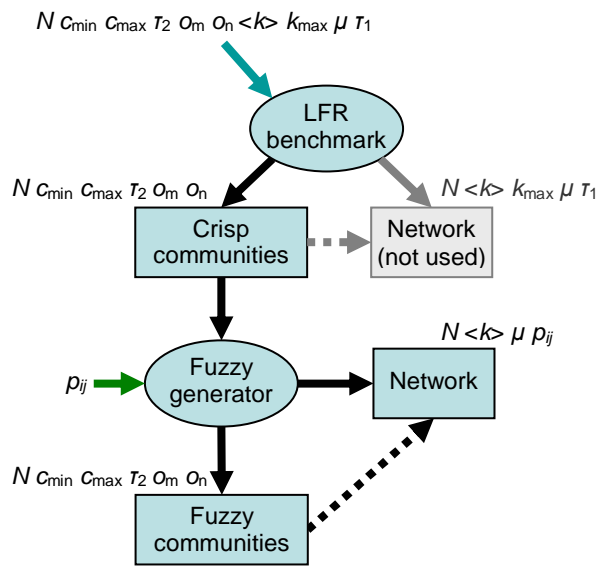

**Figure 1.** Fuzzy network generator.

*4.2. Crisp algorithms*

In this section we evaluate several "crisp" community detection algorithms on networks with both crisp and fuzzy overlapping, as defined in section 3.1. If the network contains fuzzy overlapping, the crisp partition found is treated as an approximation to a fuzzy solution and compared with the fuzzy partition used to construct the network, using the Fuzzy Rand Index. If the network contains crisp overlapping, the solution (a crisp partition) is compared with the crisp partition used to construct the network. For consistency we also use the Fuzzy Rand Index for these.

The algorithms evaluated are CFinder [4], CONGA [9], LFM [11], COPRA [14], GCE [13], and EAGLE [18]. For CONGA, we specify the correct number of communities as parameter. CFinder and COPRA each have a small integer parameter: we use $k$=4 and $v$=4, respectively. For the other algorithms we use only the default parameters.

We run each algorithm on networks with two values of mixing ($\mu \in \{0.1, 0.3\}$) and average degree ($\langle k \rangle \in \{12, 24\}$), and vary the fraction of overlapping vertices ($o_n/N$) from 0.02 to 1. The other parameters are $N$=500, $k_{max}$=$\langle k \rangle \times 2.5$, $c_{min}$=$\langle k \rangle \times 2/3$, $c_{max}$=$c_{min} \times 5$, $\tau_2$=−1, $\tau_1$=−2, $o_m$=2. All results are averaged over 100 networks with each set of parameters. Figure 2 shows the Fuzzy Rand Index of the solutions for fuzzy and crisp networks, with all four sets of parameters.

Our fuzzy and crisp networks differ in two respects. First, when the fraction of overlapping vertices is less than 1, the expected degree of a two-community vertex is greater than that of a one-community vertex in crisp networks; in fuzzy networks there is little difference. Second, even when all vertices are overlapping, each vertex belongs equally to its two communities in crisp networks, but not necessarily in fuzzy networks. This means that any algorithm that erroneously assigns a vertex to a single community can achieve a higher score on a fuzzy network than on a crisp one, by assigning each vertex to the community to which it belongs most strongly. Figure 2 shows that the performance of most methods is strongly affected by the type of overlapping. As expected, results are generally worse for crisp overlapping, except for CONGA and GCE, which are almost as effective as with fuzzy overlapping.

Performance is generally worse for higher mixing ($\mu = 0.3$) and declines as overlap increases, as expected. Most algorithms perform well on fuzzy networks with low mixing ($\mu = 0.1$). Anomalous behaviour is shown by CONGA on crisp networks: it reaches a peak when the fraction of overlapping vertices is about 0.5. This is because CONGA suffers from poor performance in the presence of mixing ($\mu > 0$): an intercommunity edge can be mistaken for overlapping, so some vertices are incorrectly placed in too many communities even when overlap is low.

We noted in section 3.2 that, for crisp networks, the results could be measured using a non-fuzzy measure such as the Omega Index or NMI, instead of the Fuzzy Rand Index. In figure 3 we plot the results of two of our partitions of crisp networks (from figure 2) using all three measures. This shows that they are all very similar.

*4.3. Fuzzy algorithms*

The MakeFuzzy method, introduced in section 3.2, allows us to obtain a non-trivial fuzzy partition from any crisp algorithm. The Fuzzy Rand Index of this fuzzy partition can be compared with that of the crisp partition (shown in figure 2) computed by the crisp algorithm itself. Figure 4 shows this comparison for two crisp algorithms: CONGA and GCE.

The results of CONGA are dramatically improved by MakeFuzzy, for both fuzzy and crisp overlapping. This is because MakeFuzzy compensates for CONGA's tendency to assign vertices to incorrect communities (noted above), by giving these occurrences a low belonging coefficient. With MakeFuzzy, CONGA gives similarly good results for both fuzzy and crisp networks.

For GCE, MakeFuzzy slightly improves the results for networks with fuzzy overlapping, successfully recovering the belonging coefficients. For networks with crisp overlapping, MakeFuzzy makes the results slightly worse. This is expected, because there is no membership information in the partition from which the network is constructed. MakeFuzzy has a similar effect on the other crisp algorithms (CFinder, LFM, COPRA, and EAGLE), so we do not show them in figure 4.

In figure 5 we compare fuzzy community detection algorithms: the six crisp algorithms extended by MakeFuzzy and two genuine fuzzy algorithms. These are Fuzzyclust [15] (with the correct number of communities as parameter) and the NMF algorithm [17] (with default parameters). All algorithms produce a fuzzy partition, which we compare with the fuzzy partition used to construct the network.

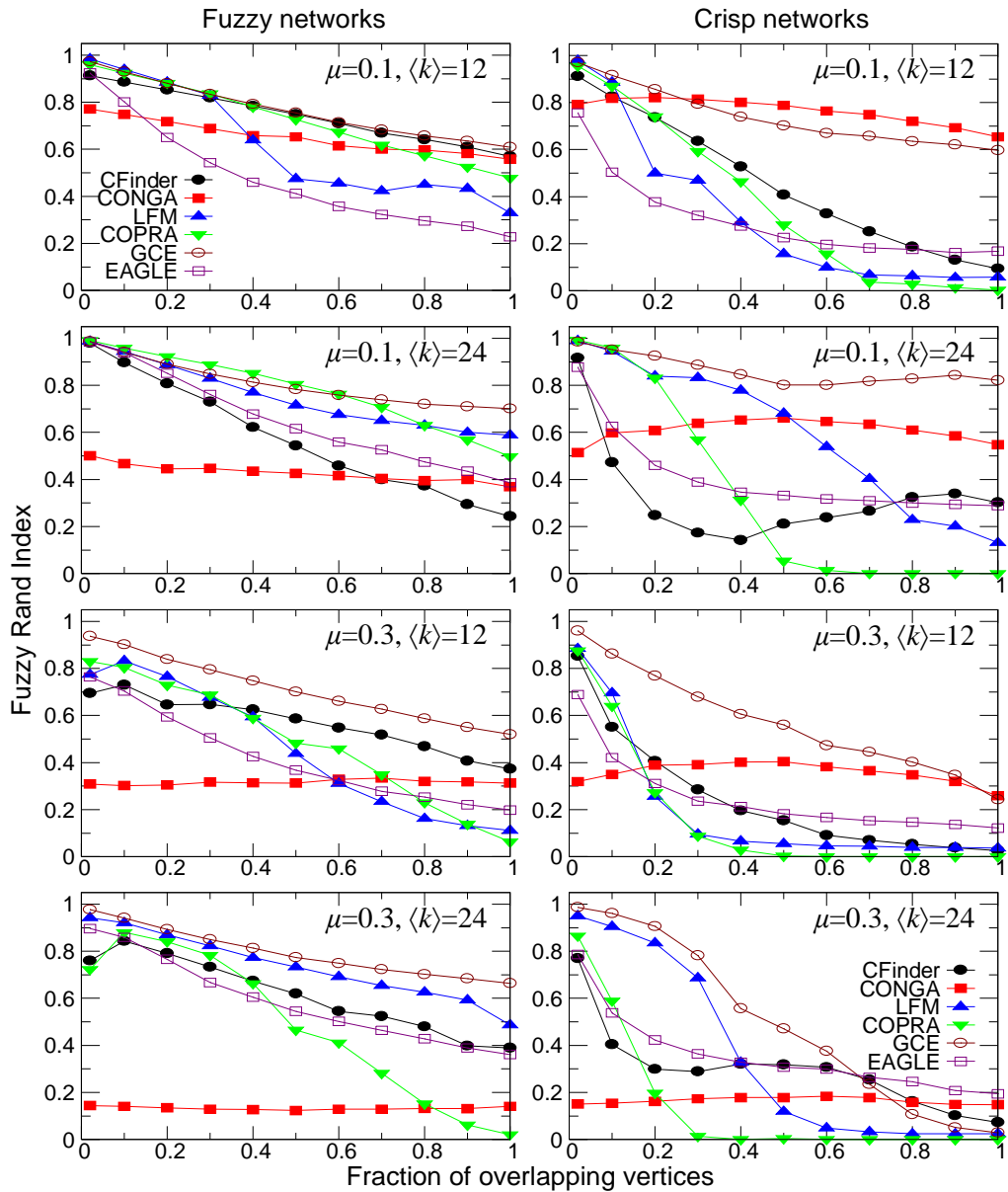

**Figure 2.** Results (Fuzzy Rand Index) of crisp algorithms (CFinder, CONGA, LFM, COPRA, GCE, and EAGLE) on networks with fuzzy (left) and crisp (right) overlapping.

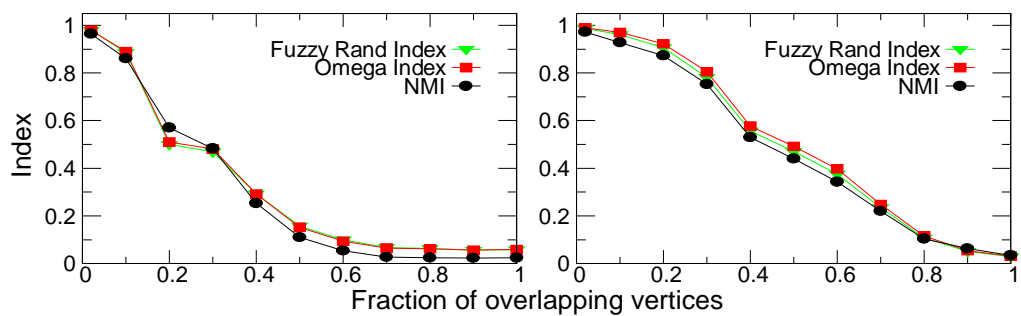

**Figure 3.** Results (Fuzzy Rand Index, Omega Index, and NMI) of some crisp algorithms on networks with crisp overlapping. Left: LFM algorithm on network with $\mu=0.1$ and $\langle k \rangle=12$. Right: GCE algorithm on network with $\mu=0.3$ and $\langle k \rangle=24$.

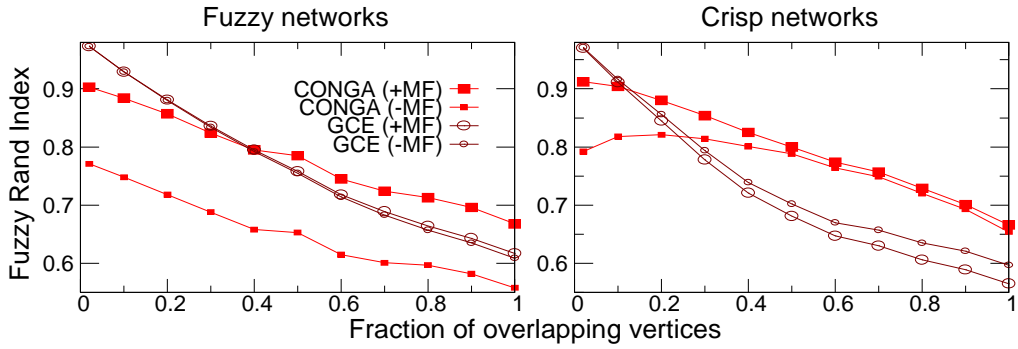

**Figure 4.** Results (Fuzzy Rand Index) of crisp algorithms CONGA and GCE with and without MakeFuzzy, on networks with fuzzy (left) and crisp (right) overlapping. Parameters: $\mu=0.1$, $\langle k \rangle=12$.

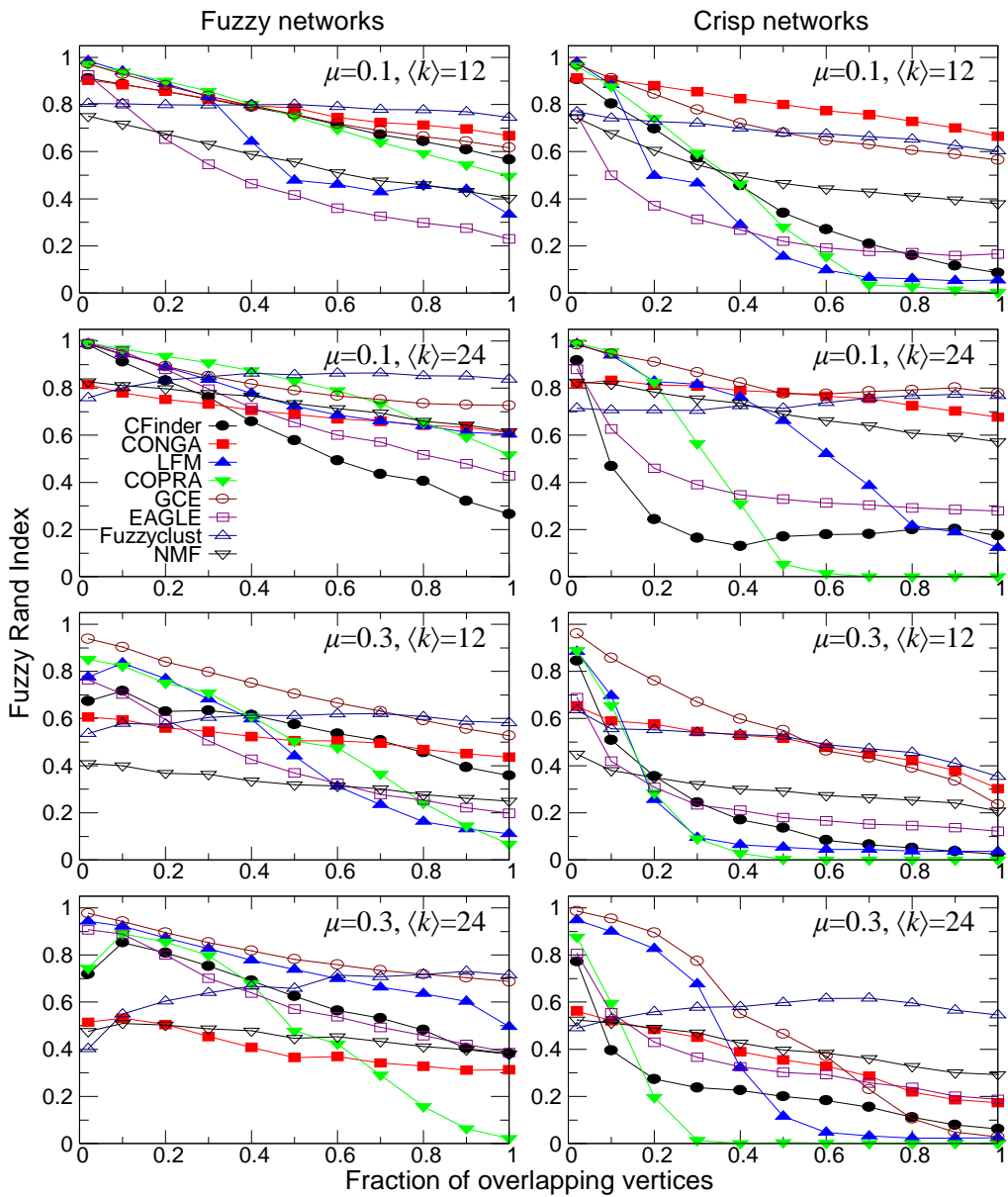

**Figure 5.** Results (Fuzzy Rand Index) of fuzzy algorithms on networks with fuzzy (left) and crisp (right) overlapping. The algorithms include genuine fuzzy algorithms (Fuzzyclust and NMF) and crisp algorithms (CFinder, CONGA, LFM, COPRA, GCE, and EAGLE) extended by MakeFuzzy.

The results confirm that, combined with MakeFuzzy, CONGA performs equally well for crisp networks and fuzzy networks. This is also generally true of GCE and both Fuzzyclust and NMF, while the other algorithms perform worse on crisp networks. Unlike all other algorithms, Fuzzyclust often performs better as overlap increases; this may be because this algorithm is given the correct number of communities and the communities are larger when overlap increases.

*4.4. Real networks*

When analysing a real network, we generally do not know whether its communities overlap fuzzily or crisply. Communities may not even overlap at all: usually, algorithms for detecting either disjoint or overlapping communities are employed without knowing whether the network's communities actually overlap. If we could detect the amount, and crispness, of overlap in a real network, it would help in choosing a suitable community detection algorithm.

Lancichinetti *et al* [33] recently proposed a method that can identify overlap in a network. This is based on the assumption that a good *disjoint* community detection algorithm will place each vertex in the same community as the greatest fraction of its neighbours. First, such an algorithm is used to find communities. Then, for each vertex, $k_{in}$, the number of neighbours that have been assigned to the same community, is measured. The *embeddedness* of $v$, $k_{in}/k$, is the fraction of $v$'s neighbours in the same community as $v$. If $k_{in}/k < 1$, assuming that there is no mixing (i.e., no intercommunity edges exist), then $v$ is assumed to belong to more than one community. If $k_{in}/k < 0.5$, $v$ must belong to at least three communities, by the same reasoning. In general, vertices in multiple communities will have low embeddedness; the lower the embeddedness, the more communities the vertex is likely to belong to.

We now extend this idea to assess the crispness of overlapping in a network. With crisp overlapping, a vertex belonging to more than one community will tend to have a higher degree than a vertex in a single community, while with fuzzy overlapping, the degree should be less affected by the number of communities. Therefore, we can measure crispness by examining the relation between a vertex's degree and the number of communities to which it belongs, estimated using embeddedness.

One problem mentioned, but not solved, by Lancichinetti *et al* [33] is that embeddedness is strongly related to the network's degree distribution. For example, the embeddedness of vertices with degree 1 is always 1, while degree-2 vertices must have embeddedness 0.5 or 1. The high frequency of low-degree vertices means that these levels of embeddedness are very common, while high-degree vertices have a wider range of possible embeddedness values. In other words, the number (and average degree) of vertices with embeddedness $e$ depends on the value of $e$ chosen.

We compensate for this as follows. We examine only a sample of vertices: those whose degree $k$ is a multiple of a small prime number, $p$. For each such $k$, we count the number of degree-$k$ vertices whose embeddedness is exactly $1/p, 2/p, …, 1$. Because this accounts for only some of the degree-$k$ vertices, we then scale each of these $p$ counts so that they sum to the total number of vertices with degree $k$. The results are used to calculate the frequency, and average degree, for each of the $p$ levels of embeddedness.

**Table 1.** Real networks used.

| Type | Name | ID | Ref. | Vertices | Edges |
|---|---|---|---|---|---|
| Social | epinions | 1 | [34] | 75879 | 405740 |
| Social | slashdot | 2 | [34] | 77360 | 469180 |
| Social | MathSciNet | 3 | [35] | 332689 | 820644 |
| Social | blogs | 4 | [9] | 3982 | 6803 |
| Social | PGP | 5 | [36] | 10680 | 24316 |
| Social | cond-mat-2003 | 6 | [37] | 27519 | 116181 |
| Biological | protein-protein | 7 | [4] | 2614 | 6379 |
| Information | google | 8 | [34] | 875713 | 4322051 |
| Information | amazon | 9 | [34] | 410236 | 2439437 |
| Information | HepTh | 10 | [34] | 27769 | 352285 |
| Communication | email-EuAll | 11 | [34] | 265009 | 364481 |
| Communication | email-Enron | 12 | [34] | 36692 | 183831 |
| Other | word_association | 13 | [4] | 7207 | 31784 |

We analyse several real networks, listed in table 1. Like Lancichinetti *et al* [33], we use the Infomap algorithm [38] to find disjoint communities. Figure 6(a) shows the embeddedness distribution of some of these. Using $p \in \{2, 3, 5\}$, we plot the frequency of vertices with embeddedness 1/2, 1/3, 2/3, 1/5, 2/5, 3/5, 4/5, and 1, normalized so that the frequency of vertices with embeddedness 1 (single-community vertices) is always 1. The plot shows that the "word_association" network has most vertices in more than one community and many in several; this is correct because this network is well known to have highly overlapping community structure. Other networks plotted appear to have less overlap, especially "amazon" and "cond-mat-2003", most of whose vertices fall into only one community. This also seems to correspond to reality; for example, "cond-mat-2003" consists of largely independent but overlapping communities of collaborating researchers.

Figure 6(b) shows how average degree varies with embeddedness for the same networks. Again, the plot is normalized so that the degree of vertices with embeddedness 1 is always 1. The "amazon" network appears clearly fuzzy: degree is unaffected by embeddedness. This seems reasonable given the nature of this network: vertices represent products and edges link co-purchased products. If a product appears in (say) two co-purchasing communities, probably because it covers two topics, there is no inherent reason why it should be co-purchased with more items, and therefore sell more copies. The other networks seem to have crisp overlapping: the degree of their vertices steadily increases as embeddedness decreases. For example, "cond-mat-2003" has much higher crispness than "amazon", despite the similar overlap. This collaboration network seems to comprise communities held together by prolific researchers, who participate equally in each of the communities that they belong to.

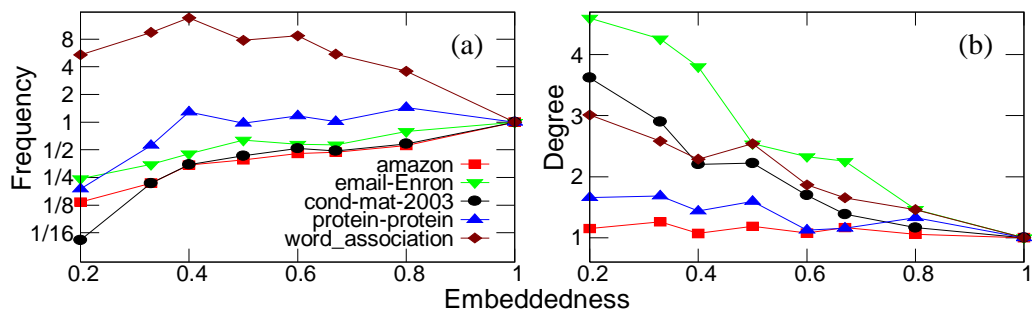

**Figure 6.** (a) Embeddedness distribution (frequency of vertices with exactly the specified embeddedness). (b) Average degree of vertices as a function of their embeddedness. All values are relative to the values for fully embedded vertices (those with all neighbours in the same community).

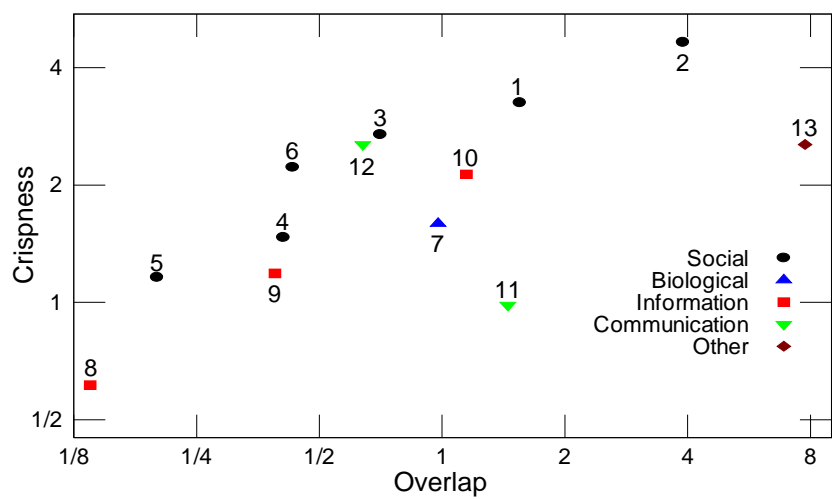

**Figure 7.** Overlap and crispness of several real networks. Overlap is the frequency, and crispness the average degree, of vertices with embeddedness 0.5 relative to those with embeddedness 1. Numbers refer to the network identifiers in table 1.

To summarize the information in figure 6, we could define *overlap* as the number of vertices with embeddedness 0.5 divided by the number with embeddedness 1 and *crispness* as the average degree of vertices with embeddedness 0.5 divided by that of vertices with embeddedness 1. Figure 7 plots these two values for all of the networks in table 1.

These results show that real networks vary widely in both overlap and crispness. It is difficult to draw any general conclusions because the sample of networks examined is small, and the results depend on the ability of the Infomap algorithm to find good disjoint partitions, which is less likely if overlap is substantial. However, it is clear that many real networks, including social networks, have substantial crispness. This suggests that our "crisp" benchmarks may be more representative of real networks than "fuzzy" ones, and that the few community detection algorithms that perform well on crisp networks may be able to handle the widest range of real networks.

## 5. Conclusions

Our main result is that, in networks with overlapping communities, the fuzziness of overlapping makes a significant difference to the ease of detecting communities. This implies that a user interested in finding overlapping communities should choose an algorithm appropriate for the type of overlap. For example, CONGA and GCE seem best suited to crisp overlapping, while many other algorithms only work well for fuzzy overlapping. It also suggests that fuzziness should be considered when overlapping community detection algorithms are benchmarked. Current benchmarks [32] feature simple forms of overlapping but do not allow the fuzziness to be varied.

Another result concerns the detection of belonging coefficients when overlapping is fuzzy. Our MakeFuzzy technique makes little improvement to the solution quality in terms of Fuzzy Rand Index, suggesting that there is a need for more special-purpose fuzzy algorithms like Fuzzyclust [15]. Nevertheless, the MakeFuzzy technique could still be useful because the motivation for recovering belonging coefficients is not only to get closer to the correct solution, but also for other purposes such as detecting roles of individuals in communities (e.g., [21]).

Our final contribution is a proposal for a method of assessing the crispness of overlapping in real networks, which we have demonstrated on a few example networks. This method could be useful in selecting a suitable algorithm to detect communities in a particular network. For example, if the network is fuzzy, there is a wider choice of effective algorithms available. Our preliminary analysis suggests that crisp overlapping is common in real networks.

An important topic for future research is to perform a more systematic analysis of crisp and fuzzy overlapping in real networks (section 4.4). Another is the design of overlapping community detection algorithms that are tailored to the different forms of overlapping. Finally, there is a need to develop alternative measures, like the Fuzzy Rand Index, for comparing fuzzy partitions and to characterize them statistically.

Our fuzzy network generator (section 4.1) will be available from http://www.cs.bris.ac.uk/~steve/networks/ .

**Acknowledgements**
I am grateful to Tamás Nepusz and Giuseppe Mangioni for discussions on fuzzy overlapping, and to them, Conrad Lee, and the anonymous referees for comments on a draft of this paper. Thanks are also due to Xiaonan Zhang, who performed some preliminary experiments on this topic in his MSc dissertation.